\documentclass[twocolumn,aps,pra,superscriptaddress,amssymb]{revtex4}
\usepackage{natbib}
\usepackage{graphicx}
\usepackage{units}
\usepackage{textcomp}
\usepackage{psfrag}
\usepackage{color}
\usepackage{amssymb}
\usepackage{amsbsy}

\begin{document}

\hyphenation{Ryd-berg}


\title{Universal scaling in a strongly interacting Rydberg gas }

\author{Robert L\"ow}
\email[Electronic address: ]{r.loew@physik.uni-stuttgart.de}
\affiliation{5. Physikalisches Institut, Universit\"{a}t
Stuttgart, Pfaffenwaldring 57, 70550 Stuttgart, Germany}
\author{Hendrik Weimer}
\affiliation{Institut f\"ur Theoretische Physik III, Universit\"at Stuttgart, 70550 Stuttgart, Germany}
\author{Ulrich Raitzsch}
\affiliation{Department of Physics, Durham University, Durham DH1 3LE, U.K. }
\author{Rolf Heidemann}
\author{Vera Bendkowsky}
\author{Bj\"orn Butscher}
\affiliation{5. Physikalisches Institut, Universit\"{a}t
Stuttgart, Pfaffenwaldring 57, 70550 Stuttgart, Germany}
\author{Hans Peter B\"uchler}
\affiliation{Institut f\"ur Theoretische Physik III, Universit\"at Stuttgart, 70550 Stuttgart, Germany}
\author{Tilman Pfau}
\email[Electronic address: ]{t.pfau@physik.uni-stuttgart.de}
\affiliation{5. Physikalisches Institut, Universit\"{a}t
Stuttgart, Pfaffenwaldring 57, 70550 Stuttgart, Germany}

\date{\today}

\begin{abstract}
We study a gas of ultracold atoms
resonantly driven into a strongly interacting Rydberg state. The long distance
behavior of the spatially frozen effective pseudospin system is determined by a set of dimensionless parameters, and
we find that the experimental data exhibits algebraic scaling laws for the
excitation dynamics and the saturation of Rydberg excitation. Mean field
calculations as well as numerical simulations provide an excellent agreement
with the experimental finding, and are evidence
for universality in a strongly interacting frozen Rydberg gas.
\end{abstract}


\maketitle

\section{Introduction}

The concept of universality appears in many different fields of physics
\cite{Huang1987}, biology \cite{West2004}, economics \cite{Stanley1996} and various
other systems.  It allows to describe the behavior of a system without actually
knowing all the microscopic details of its state. A particular class of
universal scaling behavior can be found close to second order phase
transitions. The characterization of the corresponding critical points in terms of
universality classes \cite{Fisher1998} has become crucial for the understanding
of classical as well as quantum phase transitions. Quantum degenerate gases can serve
as a well controlled model system for the
exploration of universal scaling behavior and quantum phase transitions
\cite{Greiner2002} in strongly interacting cold atomic systems.  Here, we show
that the experimental data supports the appearance of universal scaling in
ultracold Rydberg gases, which is in agreement with the recently predicted
existence of a quantum critical point \cite{Weimer2008a}.\\
The key ingredients of the described experiments are the combination of a
Rydberg gas  in the `frozen' regime \cite{Anderson1998} with strong interactions
among the Rydberg atoms \cite{Heidemann2007}, and the ability to coherently
drive the system \cite{Raitzsch2008} as a pseudo spin. There exists a variety of interaction
mechanisms among Rydberg atoms giving rise to blockade phenomena, which are intensively
studied \cite{Tong2004, Vogt2006,Heidemann2007} experimentally.  Recently several groups also
focused on the coherent properties of frozen Rydberg gases in the regime of weak
\cite{Mohapatra2007,Johnson2008,Reetz-Lamour2008}, as well as strong interactions
\cite{Raitzsch2008,Urban2009,Gaetan2009,Weatherill2008,Raitzsch2009}. This unique combination of strong
interactions with long coherence times led to various proposals for quantum
information processing using Rydberg atoms
\cite{Jaksch2000,Lukin2001,Brion2007,Muller2009}.\\
\begin{figure}
\includegraphics[width=8.5cm]{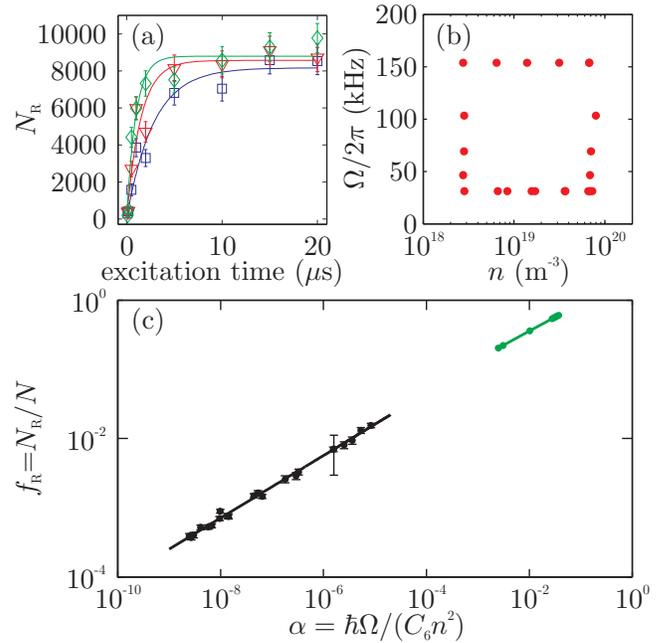}
\caption{Universal scaling of the Rydberg fraction in the saturated excitation regime (a) The saturated Rydberg excitation is obtained with a laser coupling strength  of $\Omega$=$2\pi\times$154 kHz in
dense ultracold atomic clouds with densities $n= $ [\unit[3.2$\times 10^{19}$] (\textcolor{green}{$\diamond$}),
\unit[6.6$\times 10^{18}$] (\textcolor{red}{$\triangledown$}),
\unit[2.8$\times 10^{18}$] (\textcolor{blue}{$\square$})] {m$^{-3}$} \cite{Heidemann2007}. (b) Scanned
parameter space for the individual excitation curves depicted in the  $n$ - $\Omega$ plane.
(c) Saturated Rydberg fraction $f_{R}$ as a function of the
dimensionless parameter $\alpha = \hbar \Omega/ C_{6}n^2$  for a three dimensional configuration
($\blacksquare$) and numerical simulations (\textcolor{green}{\Large $\bullet$}). The experimental and numerical data are fitted
(solid lines) to power laws of the form $f_R\sim\alpha^{1/\delta}$
from which the critical exponents $1/\delta=0.45\pm0.01$ (exp.) and $1/\delta=0.404$ (num.) are extracted.
}
\label{fig1}
\end{figure}
In this letter, we apply the theoretical framework of a quantum
critical behavior in strongly interacting Rydberg gases
\cite{Weimer2008a} to experimental data (see Fig.~\ref{fig1}a), which
has been previously analyzed with respect to coherent and collective
excitation of Rydberg atoms in the strong blockade regime
\cite{Heidemann2007}. The relevant parameters of the experiment are
the density of particles $n$, the coupling strength of the driving
laser field $\Omega$ (see Fig.~\ref{fig1}b), its detuning from
resonance $\delta_L$, and the interaction strength among the Rydberg
states determined in our case by the van der Waals constant $C_6$.  On
resonance ($\delta_L=0$), these parameters can be merged into a single
dimensionless parameter $\alpha=\hbar \Omega/C_6n^2$.  We find, that
all experimental data taken from \cite{Heidemann2007} collapse to a
simple power law as a function of this parameter $\alpha$, see
Fig.~\ref{fig1}c), which is in agreement with the predicted universal
scaling behavior. This implies that the interpretation
in terms of a continuous quantum phase transition \cite{Weimer2008a}
is supported by experimental data, and thus provides a firmer
theoretical foundation compared to previous scaling models
\cite{Heidemann2007}. While the scaling models essentially depend on
two parameters (Rabi frequency and ground state density), there is
only one single parameter in the universal scaling function, which is the length scale diverging at the critical point. We also show that these scaling models can be rigorously derived from the mean-field approach given in \cite{Weimer2008a}. \\
\begin{figure}
\includegraphics[width=8.5cm]{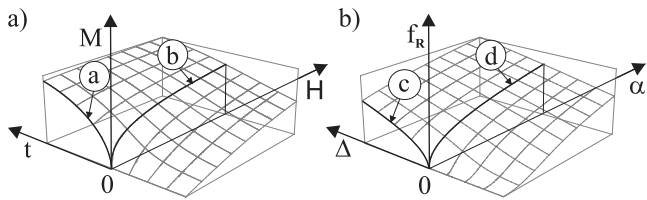}
\caption{Equation of state of a ferromagnet in comparison to a strongly interacting Rydberg gas a) Magnetization $M$ of a classical ferromagnetic Ising model: The
system exhibits a critical point at $t=1-T/T_{c}=0$ and $H=0$.  The
magnetization $M$ is given by a power laws \textcircled{\textsf{a}} $M\sim
t^\beta$ and \textcircled{\textsf{b}} $M\sim H^{1/\delta}$. The exponents have
been calculated with the Ising model to be $\beta=0.31$ (measured: 0.32-0.39) and
$\delta=5$ (measured: 4-5) \cite{Huang1987}. b) Excited state fraction $f_R$ of a strongly interacting Rydberg gas.
The universal scaling laws are given by power laws of the rescaled detuning
$\Delta$ of the driving field \textcircled{\textsf{c}} $f_{R}(\alpha=0)\sim \Delta^\beta$ and
the rescaled coupling strength $\alpha$ as depicted by \textcircled{\textsf{d}} $f_{R}(\Delta=0)\sim
\alpha^{1/\delta}$. Within mean-field theory the plotted parameter $f_R$ is given  for
a three dimensional cloud with van der Waals interaction by
$\alpha = f_R^{\delta} | 1-\Delta/f_R^{1/\beta}|$ with $\delta= 5/2$ and
$\beta = 1/2$.}
\label{fig2}
\end{figure}

\section{Universal scaling theory}

Textbooks on statistical mechanics \cite{Huang1987} often introduce universal
scaling as a critical phenomenon, which can be found near the critical point of
a second order phase transition. 
When approaching the critical point the system becomes
scale invariant. This means that microscopic details of the system
become irrelevant, and the macroscopic behavior is dominated by its
long-range physics, associated with a diverging length scale
$\xi$. Mathematically, a function $f(s)$ is called scale-invariant,
if $f(\lambda s)=c f(s)$. Using a series expansion one can see that
solutions to this equation are given by power laws of the form
$f(s)\propto s^{\nu}$. Hence, near the critical point all
observables can be described by power laws of the diverging scale
$\xi$. Since the critical properties are dominated by long-range
physics many different systems show the same critical behavior. This
leads to the classification of critical exponents in terms of
universality classes, which are determined by the spatial dimension,
the symmetries of the Hamiltonian and the long-range behavior of the
interactions.

Phase transitions occur when the free energy of a
system shows nonanalytic behavior. In classical systems this is
always related to a change in temperature. However, in quantum
systems at $T=0$ there is a another possibility: the ground state
energy can become nonanalytic in the case of an avoided crossing
with vanishing gap or an actual level crossing
\cite{Sachdev1999}. Analogously to their classical counterparts one
can classify quantum phase transitions into first, second, and
inifinite order transitions.


\subsection{Universal scaling in a Ferromagnet}

The most prominent example of a second order phase transition is a ferromagnet close to the Curie
temperature $T_C$. For $T>T_C$ the system is completely demagnetized and
rotationally invariant.  Lowering the temperature below $T_C$, the system enters
the ferromagentic phase characterized by a finite magnetization $M$, which
breaks the rotational symmetry.  The appearance of the magnetization can be
described in terms of an order parameter.  Above $T_C$ the magnetization $M$ is
simply zero, and exhibits a linear magnetization in the presence of an external
magnetic field $H$. The equation of state close to the critical point at $T_{c}$
is shown in Fig.~\ref{fig2}a) in terms of the reduced temperature
$t=1-T/T_C$ and the external magnetic field $H$. Then the system is described in
terms of universal scaling laws of the form $M\sim t^\beta$ for $H=0$ and $M\sim
H^{1/\delta}$ for $t=0$, indicated by the curves \textcircled{\textsf{a}} and
\textcircled{\textsf{b}}, while the diverging
length satisfies $\xi \sim 1/t^\nu$.

\subsection{Universal scaling in a strongly interacting Rydberg gas}

Now we want to draw the analogy of the
magnet close to the Curie point to a strongly interacting Rydberg gas driven by
a laser field. The frozen atomic gas is assumed to consist of an ensemble of spatially fixed
pseudospins with two electronic states, one being the ground
state and the other a highly excited Rydberg state. The coupling $\Omega$
between the two states is achieved by a monochromatic light field with a
detuning $\delta_L$ with respect to the energy splitting of the two states.  The
Rydberg states interact strongly with  a general interaction potential
$C_p/r^p$, which in the present experimental situation is dominated by the van
der Waals interaction with $p=6$. The corresponding $N$-particle Hamiltonian then
reads
\begin{equation}\label{eq.Ham}
H=-\frac{\hbar \delta_L}{2} \sum_i \sigma_z^{(i)}+\frac{\hbar \Omega}{2} \sum_i
\sigma_x^{(i)}+C_p \sum_{j<i}
\frac{P_{ee}^{(i)}P_{ee}^{(j)}}{|\textbf{r}_{i}-\textbf{r}_{j}|^p},
\end{equation}
where the $\sigma^{(i)}_{x,y,z}$ are Pauli matrices, $\textbf{r}_{i}$ the position of the
atom $i$ and $P_{ee}^{(i)}=(1+\sigma_z^{(i)})/2$ is the projector on the excited
Rydberg state. It has been recently shown \cite{Weimer2008a}, that this
Hamiltonian features a quantum critical point at $\Omega=\delta=0$ for $p>d$,
where $d$ is the dimensionality of the system. In contrast to the above
classical example, the critical point appears even at zero temperature by
varying the detuning $\delta_{L}$. Consequently, the role of the reduced
temperature $t$ is now taken by the dimensionless detuning $\Delta =
\hbar\delta_{L}/E_{c}$ with the characteristic energy $E_c=C_p n^{p/d}$.  For
$\Delta<0$, all atoms remain in the ground state for $\Omega\rightarrow 0$, while for
$\Delta>0$ a finite number $N_{R}$ of atoms are excited into the Rydberg state.
The fraction of excited Rydberg atoms $f_{R}=N_{R}/N$ then plays the analog
role of the magnetization $M$ in the example of a ferromagnetic
phase transition discussed above. In analogy, we can draw the fraction of excited Rydberg
atoms $f_{R}$ close to the critical point as a function of the detuning
$\Delta$ and the parameter $\alpha=\hbar\Omega/E_{c}$, see
Fig.~\ref{fig2}b). Note, that the continuous behavior of $f_{R}$ at
the critical point is a special property of the long-distance behavior of the
interaction potential. The coherence length $\xi$ is determined by the
characteristic length scale of the
correlation function $\langle (P_{ee}^{(i)}-f_{R})(P_{ee}^{(j)}-f_{R})\rangle$, which gives
rise to a diverging length scale close to the critical point. This quantity
corresponds to the blockade radius in the system, i.e., the
radius around a Rydberg atom, up to which an additional Rydberg excitation is
strongly suppressed.\\

It remains to identify the critical exponents $\beta$ and $1/\delta$ for the
universal scaling of the observable parameter $f_R\sim \Delta^\beta$
and $f_R\sim \alpha^{1/\delta}$. For $\alpha=0$ the Hamiltonian
(\ref{eq.Ham}) is classical and by minimizing its energy one obtains
$\beta=d/p$. The correlation length $\xi$ is determined by the
averaged spacing between the Rydberg atoms via $\xi \sim a / f_R^{1/d}$ (here,
$a^{-1}=\sqrt[d]{n}$ denotes the averaged interparticle distance of a $d$-dimensional system).
To obtain a value for $1/\delta$, it is useful to take a closer look at the
excitation dynamics of a strongly interacting Rydberg system ($\alpha\ll 1$)
driven at resonance ($\Delta=0$), which also complies with the experimental
situation \cite{Heidemann2007}. When $\alpha$ approaches zero, the system enters the strongly
blocked regime with $f_{R} \ll 1$.  Within the blockade radius $\xi$ only
one excitation is shared over a large number of atoms $N_b \sim n \xi^d$
resulting in a collective state
$|\psi_e\rangle=\frac{1}{\sqrt{N_b}}\sum_{i=1}^{N_b}|g_1,g_2,g_3,...,e_i,...g_{N_b}\rangle$,
which shows an accelerated temporal evolution with a collective Rabi
frequency $\sqrt{N_b}\Omega$ \cite{Heidemann2007,Hernandez2008,Gaetan2009}. In this so called superatom
model the blockade radius can be evaluated by equating the interaction energy
with the collective coupling strength as $C_p/\xi^p=\sqrt{N_b}\Omega$. Note,
that $\xi$ diverges for $\alpha \rightarrow 0$. This
procedure determines $1/\delta$ to be $1/\delta=2 d/(2 p+d)$ and the case for
$d=3$ and $p=6$ is carried out in some more detail in \cite{Heidemann2008}.
The same result for $1/\delta$ is obtained using standard mean-field theory
\cite{Weimer2008a}.
In addition, the mean-field solution leads to a description of the full
behavior of $f_R$ in terms of a general scaling function
\begin{equation}
  f_R = \alpha^{2 d/(2 p +d)} \: \: \chi\left(\frac{\Delta}{\alpha^{2 p/(2 p +
d)}}\right).
\end{equation}
This is the inverse of the exact mean field result
$\alpha = f_R^{\delta} | 1-\Delta/f_R^{1/\beta}|$ shown in Fig.~\ref{fig2}b).
As the system turns classical in the limit $\alpha \rightarrow 0$, the behavior
of the unknown function $\chi(y)$ in the limit $y \rightarrow \pm \infty$ can
exactly be determined to be
$\chi(y) \sim y^{d/p}$ and $\chi(y) = 1/y^2$, respectively. It is important to note
that the mean-field result for $\delta$ is the only consistent scaling
exponent with these limits. This indicates that the classical behavior of the
system at $\alpha = 0$ fixes the exponent $\delta$ to its mean field value
even for low dimensions. In contrast to a classical critical point,
for a quantum critical point the dynamical behavior is coupled to the
static properties via the dynamical critical exponent $z$, i.e.
$\tau \sim \xi^z$
where $\tau$ describes the characteristic time scale close to the
critical point.  Here, we find the
dynamical critical exponent to be $z=p$, which implies that the
relaxation is domitated by the frequency
$ \sqrt{N_{b}} \Omega$ in agreement with the above superatom picture. \\

\section{Experiment}

\subsection{Universal scaling in an inhomogeneous sample}

In the actual experiment, the atoms are well described as a thermal gas trapped by a
harmonic potential. Then, the  three dimensional density distribution of the $N$
ground state atoms has a Gaussian shape with radii given by the standard
deviations $\sigma_{x,y,z}=\sqrt{k_B T/2m\omega_{x,y,z}}$, which are
determind by the trapping frequencies $\omega_{x,y,z}$, the mass $m$, the
temperature $T$ of the cloud and Boltzmanns constant $k_B$
\begin{equation}\label{eq.densitydist}
n(\textbf{r})=\frac{N}{(2\pi)^{3/2} \sigma_x \sigma_y \sigma_z} \exp
\left(-\frac{x^2}{2 \sigma_x^2}-\frac{y^2}{2 \sigma_y^2}-\frac{z^2}{2
\sigma_z^2}\right).
\end{equation}
Note, that the temperature $T$
associated with the kinetic energy of the atoms is decoupled from the dynamics
of the Rydberg excitations in the frozen  Rydberg gas. Within the local
density approximation, we can describe the properties of the system by a local
parameter $\alpha({\bf r}) = \hbar \Omega/C_{6} n({\bf r})^2$ and the total
Rydberg fraction $f_R$ is given by ($d=3,\, p=6$)
\begin{equation}\label{eq.rydfracint}
f_R=\frac{1}{N}\int dr^3 f_R(\textbf{r})n(\textbf{r}) \sim
\frac{1}{(-2/\delta+1)^{3/2}}\alpha^{1/\delta}.
\end{equation}
Here, $\alpha$ is the peak value in the trap center $\alpha=\hbar \Omega/C_6
n(\textbf{0})^2$. Consequently, we find that the critical
exponent $\delta$ is not modified by the harmonic trapping potential within the
local density approximation, and reduces to the value given in the thermodynamic
limit.\\
\begin{figure}
\includegraphics[width=8.5cm]{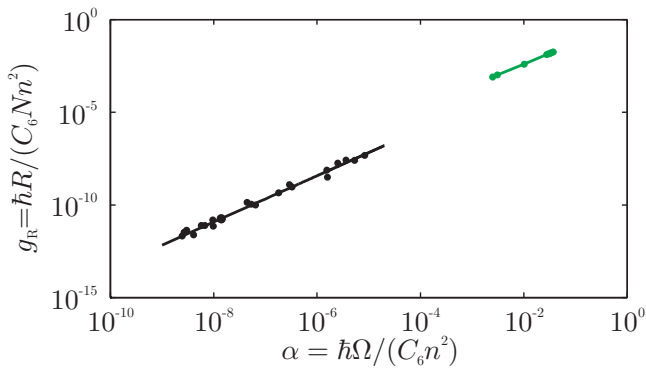}
\caption{Universal scaling behavior of the excitation rate. The rescaled excitation rate $g_{R}$ for
a three dimensional density distribution is shown for experimental data ($\blacksquare$) and the
corresponding numerical simulation (\textcolor{green}{\Large $\bullet$})).
A linear fit to a power law $g_{R}\sim\alpha^{\gamma}$ results in a critical exponent of
$\gamma=1.25\pm0.03$ (exp) and 1.15 (num).} \label{fig3}
\end{figure}

\subsection{Experimental setup and procedure}

A detailed description of the experimental setup can be found in
\cite{Heidemann2007,Low2007} and here only a rough outline of the
experimental procedure is given. First we prepare a magnetically trapped cloud
of Rubidium atoms spin polarized in the $5S_{1/2}$ state. The atomic
cloud has a temperature of 3.4 $\mu$K and peak densities $n_0$ close to
$10^{20}$ m$^{-3}$. This corresponds to phase-space densities below quantum
degeneracy to avoid a bimodal density distribution \cite{Heidemann2008}
of a Bose-Einstein condensate. To
alter the atomic density without changing the size of the cloud we use
a Landau-Zener sweep technique \cite{Raitzsch2008} to alter the number of trapped atoms from
$N=1.5\times 10^7$ to $N=5\times 10^5$. In this process
the temperature remains unchanged and consequently the physical dimensions of the cloud.
With the known harmonic oscillator
potential of the Ioffe-Pritchard type trap, all parameters of the atomic
clouds are known. The excitation to the $43S_{1/2}$ Rydberg state is done with a
resonant two-photon transition via the $5P_{3/2}$ state. To avoid population of
the intermediate $5P_{3/2}$ state the light is detuned with respect to this
state by $\delta_L$=$2\pi\times$478 MHz to the blue. The coupling strength $\Omega$ of this effective
two-level system is altered from $2\pi\times$31 kHz to
$2\pi\times$154 kHz by adjusting the laser intensity. The excitation dynamics are investigated by a variation of
the excitation time from 100 ns to 20 $\mu$s, which is short compared to the excited state lifetime of 100 $\mu$s.
After excitation the Rydberg atoms are field ionized and the emerging ions are detected with the help of a
multi-channel plate. The resulting strongly blockaded excitation dynamics is determined by the strong repulsive interaction
among Rydberg states, which is in our case given by the isotropic van der Waals interaction with $C_6=-1.7\times 10^{19}$ a.u.
for the 43S state.\\


\subsection{Experimental results}

In the experiment $\alpha$ is changed non-adiabatically by switching on abruptly
the coupling laser field $\Omega$. Then, the number of excited Rydberg atoms
$N_{R}(t)$ undergoes a dynamical evolution, which saturates in
$N_{R}$ as shown in Fig. \ref{fig1}a). The inital increase in the number of Rydberg atoms is well described
by a rate $R$, and this relaxation time is experimentally deduced by a fit of
the time-evolution of the Rydberg excitation $N_{R}(t)$ by an exponential
saturation function $N_{R}(t)=N_{R}(1 - e^{-R \: t/N_{R}})$,
which allows to extract both the initial excitation rate $R$ and the saturation
level $N_{R}$. In the strongly blocked regime $(\alpha\ll 1)$ the rate $R$ is
determined by the collective Rabi frequency $\sqrt{N_b}\Omega$ and the
saturation level $N_{R}$, which is close to that of the ground state of the system.
Previously, we have examined the data and its dependence
on the ground state density $n$ and the Rabi frequency $\Omega$ using a general
expression  $R \sim n^{\kappa_{R}} \Omega^{\lambda_{R}}$ and $N_{R} \sim
n^{\kappa_{N}} \Omega^{\lambda_{N}}$ \cite{Heidemann2007}. The scaling behavior
for a variation of the ground state density $n$ and the coupling strength
$\Omega$ gave a strong evidence for a coherent collective excitation dynamics in
the strong blockade regime.\\
\begin{table}
  \caption{Comparison of the different results for the critical exponents
$\gamma$ and $1/\delta$ for a three dimensional
as well a one dimensional density distribution. The theoretical values
are given in Eq.~(\ref{eq.ffsat1}) and Eq.~(\ref{eq.ffsat2}). The
results of the numerical simulation have been achieved by integrating the
Hamiltonian (\ref{eq.Ham}) for up to 100 particles \cite{Weimer2008a}. The
experimental results are obtained by fitting power laws to the data shown in
Fig.~\ref{fig1} and Fig.~\ref{fig3}.}
  \label{ResTable}
    \begin{ruledtabular}
    \begin{tabular}{l|lr|l}
    & $\boldsymbol{\gamma}\hspace{0.4cm} (g_R\sim\alpha^\gamma)$ & & $\textbf{1/}\boldsymbol{\delta} \hspace{0.2cm}(f_R\sim\alpha^{1/\delta})$\\
    \hline
    experiment \hspace{0.8cm} \bf{[1d]} &  $1.08 \pm 0.01$ &  &  $0.16 \pm 0.01$ \\
theory  & $14/13\approx 1.08$ &\hspace{10pt}& $2/13\approx 0.15$  \\
numerical simulation  \,\, &  1.06 &   & 0.150
\cite{Weimer2008a}   \\
\hline
experiment \hspace{0.8cm} \bf{[3d]}
 &  $1.25 \pm 0.03$ &  &  $0.45 \pm 0.01$ \\
theory  & $6/5=1.2$ &
& $2/5= 0.4$  \\
numerical simulation  \,\, & 1.15 &    &   0.404
\cite{Weimer2008a} \\
    \end{tabular}
    \end{ruledtabular}
\end{table}
In the following, we analyze these results in terms of a universal scaling
behavior at resonance $(\Delta=0)$ by rescaling the measured quantities to a
dimensionless rate $g_{R}$ and and a dimensionless saturation level $f_{R}$
as
\begin{eqnarray}
\label{eq.ffsat1}
g_R&=&\frac{\hbar R}{N C_p n^{p/d}}\sim\alpha^{\gamma}
\hspace{30pt}  \gamma = \frac{2( p+  d)}{2 p+d}, \\
\label{eq.ffsat2}
f_{R}&=&\frac{N_{R}}{N}\sim \alpha^{1/\delta}
\hspace{49pt}  \delta =\frac{2 p+d}{2 d}.
\end{eqnarray}
First, we would like to point out,
that the data collapse in to a
algebraic relations as shown Fig. \ref{fig1}c) and Fig. \ref{fig3}) is in agreement with the predicted scaling laws.
It is worth to mention that the numerical simulations based on only $10^2$ pseudospins
scales up to the experimental situation with atom numbers of up to $10^7$. For a more quantitative analysis of the experimental
data, it is important to point out that the radial Gaussian radius of the cigar
shaped cloud is $\sigma_{x,y}=8.6\, \mu {\rm m}$. This width is comparable to the
blockade radius between two Rydberg atoms of roughly $5\,\mu {\rm m}$ and places by this
the geometry in a crossover regime between one dimension and three dimensions.
Therefore we analyze the data wether the experimental setup is better
described in one dimension with a line density ($\alpha=\hbar \Omega/C_6n^6$) or in three dimensions ($\alpha=\hbar \Omega/C_6n^2$).
Fitting the observed power laws of the form $\ln g_R\sim \gamma
\,\,\ln \alpha+c_g$ and $\ln f_{R}\sim 1/\delta \,\,\ln \alpha+c_f$
we extracted the individual exponents. The results are summarized in table
\ref{ResTable} and  compared to the expected theoretical values as well
as to numerical simulations; the procedure for the numerical analysis is described in detail
in \cite{Weimer2008a}. The relatively small error bars of the
fitted exponents $1/\delta$ show the excellent agreement with power laws
over a large range in $\alpha$. Nevertheless, the dimensionality can not unambiguously
assigned, although there is a better agreement for the one dimensional case. Here the exponent is
dominated by the variation in density, which scales to the $6^{th}$ power.
In future experiments it might be possible to create better
defined dimensionality by adjusting the shape of the atomic cloud and/or the
strength of the interaction by choosing adequate Rydberg states.\\
\section{Discussion and outlook}

In summary we have shown that the experimental results presented in
\cite{Heidemann2007} can be described with universal
scaling theories. This result confirms that the description with the effective
spin Hamiltonian given in this article is correct to a large extent. Therefore
the observation of a quantum critical point in this system should be within
experimental reach. This could be done by measuring the excited Rydberg fraction when
approaching the quantum critical point adiabatically starting from a
non-critical region in the $\alpha-\Delta$ parameter space.  Another way would
be the observation of a crystalline correlation function of the excited Rydberg
atoms by either a spatial dependent observation of the Rydberg atoms or in the
Fourier space by a Laue diffraction experiment with a four wave mixing technique
\cite{Brekke2008}. The measurement of the critical exponent is not in complete
accordance with a one dimensional or three dimensional situation, which is most likely due to
finite size effects. The general
form of the scaling exponents allows also to apply the simple model to dipolar
systems, which are widely realized in frozen Rydberg gases and are also feasible
in the context of ultra-cold dipolar molecules \cite{Ni2008}.

\begin{acknowledgments}
We thank the Deutsche
Forschunggemeinschaft for financial support within the SFB/TRR21 and by
Pf381/4-1 as well the Landesstiftung Baden W\"urttemberg. B.B. acknowledges support
from the Carl Zeiss Stiftung.
\end{acknowledgments}


\begin{thebibliography}{10}

\bibitem{Huang1987}
K.~Huang,
\newblock {\em Statistical Mechanics} (John Wiley and Sons, New York, 1987).

\bibitem{West2004}
G.~B. West and J.~H. Brown,
\newblock Phys. Today {\bf 57}, 36 (2004).

\bibitem{Stanley1996}
M.~H.~R. Stanley, L.~A.~N. Amaral, S.~V. Buldyrev, S.~Havlin, H.~Leschhorn,
  P.~Maass, M.~A. Salinger, and H.~E. Stanley,
\newblock Nature {\bf 379}, 804 (1996).

\bibitem{Fisher1998}
M.~E. Fisher,
\newblock Rev. Mod. Phys. {\bf 70}, 653 (1998).

\bibitem{Greiner2002}
M.~Greiner, O.~Mandel, T.~W. Esslinger, T.~H{\"a}nsch, and I.~Bloch,
\newblock Nature {\bf 415}, 39 (2002).

\bibitem{Weimer2008a}
H.~Weimer, R.~L{\"o}w, T.~Pfau, and H.~P. B{\"u}chler,
\newblock Phys. Rev. Lett. {\bf 101}, 250601 (2008).

\bibitem{Anderson1998}
W.~R. Anderson, J.~R. Veale, and T.~F. Gallagher,
\newblock \prl {\bf 80}, 249 (1998).

\bibitem{Heidemann2007}
R.~Heidemann, U.~Raitzsch, V.~Bendkowsky, B.~Butscher, R.~L\"ow, L.~Santos, and
  T.~Pfau,
\newblock Phys. Rev. Lett. {\bf 99}, 163601 (2007).

\bibitem{Raitzsch2008}
U.~Raitzsch, V.~Bendkowsky, R.~Heidemann, B.~Butscher, R.~L\"{o}w, and T.~Pfau,
\newblock Phys. Rev. Lett. {\bf 100}, 013002 (2008).

\bibitem{Tong2004}
D.~Tong, S.~M. Farooqi, J.~Stanojevic, S.~Krishnan, Y.~P. Zhang,
  R.~C\^{o}t\'{e}, E.~E. Eyler, and P.~L. Gould,
\newblock \prl {\bf 93}, 063001 (2004).

\bibitem{Vogt2006}
T.~{Vogt}, M.~{Viteau}, J.~{Zhao}, A.~{Chotia}, D.~{Comparat}, and P.~{Pillet},
\newblock \prl {\bf 97}, 083003 (2006).

\bibitem{Mohapatra2007}
A.~K. Mohapatra, T.~R. Jackson, and C.~S. Adams,
\newblock Phys. Rev. Lett. {\bf 98}, 113003 (2007).

\bibitem{Johnson2008}
T.~A. Johnson, E.~Urban, T.~Henage, L.~Isenhower, D.~D. Yavuz, T.~G. Walker,
  and M.~Saffman,
\newblock Phys. Rev. Lett. {\bf 100}, 113003 (2008).

\bibitem{Reetz-Lamour2008}
M.~Reetz-Lamour, J.~Deiglmayr, T.~Amthor, and M.~Weidem\"{u}ller,
\newblock New J. Phys. {\bf 10}, 045026 (2008).

\bibitem{Urban2009}
E.~Urban, T.~A. Johnson, T.~Henage, L.~Isenhower, D.~D. Yavuz, T.~G. Walker,
  and M.~Saffman,
\newblock Nature Phys. {\bf 5}, 110 (2009).

\bibitem{Gaetan2009}
A.~Ga\"etan, Y.~Miroshnychenko, T.~Wilk, A.~Chotia, M.~Viteau, D.~Comparat,
  P.~Pillet, A.~Browaeys, and P.~Grangier,
\newblock Nature Phys. {\bf 5}, 115 (2009).

\bibitem{Weatherill2008}
K.~J. Weatherill, J.~D. Pritchard, R.~P. Abel, M.~G. Bason, A.~K. Mohapatra,
  and C.~S. Adams,
\newblock J. Phys. B {\bf 41}, 201002 (2008).

\bibitem{Raitzsch2009}
U.~Raitzsch, R.~Heidemann, H.~Weimer, B.~Butscher, P.~Kollmann, R.~L\"ow, H.~P.
  B\"uchler, and T.~Pfau,
\newblock New J. Phys. {\bf 11}, 055014 (2009).

\bibitem{Jaksch2000}
D.~Jaksch, J.~I. Cirac, P.~Zoller, S.~L. Rolston, R.~C\^ot\'e, and M.~D. Lukin,
\newblock Phys. Rev. Lett. {\bf 85}, 2208 (2000).

\bibitem{Lukin2001}
M.~D. Lukin, M.~Fleischhauer, R.~C\^ot\'e, L.~M. Duan, D.~Jaksch, J.~I. Cirac,
  and P.~Zoller,
\newblock Phys. Rev. Lett. {\bf 87}, 037901 (2001).

\bibitem{Brion2007}
E.~Brion, K.~M{\o}lmer, and M.~Saffman,
\newblock Phys. Rev. Lett. {\bf 99}, 260501 (2007).

\bibitem{Muller2009}
M.~M\"{u}ller, I.~Lesanovsky, H.~Weimer, H.~P. B\"{u}chler, and P.~Zoller,
\newblock Phys. Rev. Lett. {\bf 102}, 170502 (2009).

\bibitem{Sachdev1999}
S.~Sachdev,
\newblock {\em Quantum Phase Transitions} (Cambridge University Press,
  Cambridge, 1999).

\bibitem{Hernandez2008}
J.~V. Hern\'{a}ndez and F.~Robicheaux,
\newblock J. Phys. B {\bf 41}, 045301 (2008).

\bibitem{Heidemann2008}
R.~Heidemann, U.~Raitzsch, V.~Bendkowsky, B.~Butscher, R.~Low, and T.~Pfau,
\newblock Phys. Rev. Lett. {\bf 100}, 033601 (2008).

\bibitem{Low2007}
R.~L\"{o}w, U.~Raitzsch, R.~Heidemann, V.~Bendkowsky, B.~Butscher,
  A.~Grabowski, and T.~Pfau,
\newblock arXiv:0706.2639  (2007).

\bibitem{Brekke2008}
E.~Brekke, J.~O. Day, and T.~G. Walker,
\newblock Phys. Rev. A {\bf 78}, 063830 (2008).

\bibitem{Ni2008}
K.-K. Ni, S.~Ospelkaus, M.~H.~G. de~Miranda, A.~Pe'er, B.~Neyenhuis, J.~J.
  Zirbel, S.~Kotochigova, P.~S. Julienne, D.~S. Jin, and J.~Ye,
\newblock Science {\bf 322}, 231 (2008).

\end{thebibliography}
\end{document}